\documentclass[preprint,amsmath,amssymb]{revtex4}

\usepackage{graphicx}
\usepackage{dcolumn}
\usepackage{bm}

\begin{document}
\bibliographystyle{naturemag}
\preprint{}

\title{Making optical atomic clocks more stable with $10^{-16}$ level laser stabilization}

\author{Y. Y. Jiang$^{1,2}$, A. D. Ludlow$^{1\ast}$, N. D. Lemke$^{1,3}$, R. W. Fox$^{1}$, J. A. Sherman$^{1}$, L.-S. Ma$^{2}$, and C. W. Oates$^{1}$ }
\affiliation{$^{1}$National Institute of Standards and Technology, 325
Broadway, Boulder, CO  80305, USA}
\affiliation{$^{2}$State Key Laboratory of Precision Spectroscopy, East China Normal University, Shanghai, China}
\affiliation{$^{3}$University of Colorado, Department of Physics, Boulder, CO  80309, USA}
\email{ludlow@boulder.nist.gov}

\addtocounter{page}{-1}
\maketitle
\pagestyle{empty}

\pagebreak
\pagestyle{plain}
\textbf{The superb precision of an atomic clock is derived from its stability. Atomic clocks based on optical (rather than microwave) frequencies are attractive because of their potential for high stability, which scales with operational frequency.  Nevertheless, optical clocks have not yet realized this vast potential, due in large part to limitations of the laser used to excite the atomic resonance.  To address this problem, we demonstrate a cavity-stabilized laser system with a reduced thermal noise floor, exhibiting a fractional frequency instability of $2 \times 10^{-16}$.  We use this laser as a stable optical source in a Yb optical lattice clock to resolve an ultranarrow 1 Hz transition linewidth.  With the stable laser source and the signal to noise ratio (S/N) afforded by the Yb optical clock, we dramatically reduce key stability limitations of the clock, and make measurements consistent with a clock instability of $5 \times 10^{-16} / \sqrt{\tau}$.}

The instability of an atomic clock at a certain averaging time is roughly given by the ratio of the transition linewidth to transition frequency ($\delta\nu/\nu_0$) divided by the measurement S/N at that time.  Optical atomic clocks based on large ensembles of atoms benefit not only from a large $\nu_0$, but also from a potentially large S/N.  This makes them strong candidates for reaching measurement instability $\leq 10^{-17}$ in 1 s. However, this stability remains far from realized. In a seminal research effort in 2001, two optical atomic clocks at NIST were directly compared with a frequency comb \cite{UdemDiddams2001}.  The instability of these clocks was shown to be at record levels, $\thicksim 4\times10^{-15}/\sqrt{\tau}$ (for measurement time $\tau$).  Research laboratories around the world have continued to refine optical atomic clocks since, reaching new levels of ultimate uncertainty, in the most extreme case $\leq 10^{-17}$ \cite{ChouHume2010}.   However, the stability of these systems has not significantly improved, so that characterizing these clocks at improved levels requires long averaging times.  In fact, as microwave clocks (with $\nu_0$ five orders of magnitude smaller) have reached their fundamental stability limit, they have achieved an instability \cite{BizeLaurent2004} only ten times worse than the best optical clock.  Similarly, large ensemble optical clocks with promising S/N have so far only demonstrated stability similar to that of optical clocks based on single ions.  As a result, systems such as optical lattice clocks based on neutral Yb and Sr remain far from their potential.

The stability is commonly limited by an imperfect clock laser, called here the local oscillator (LO).  LO frequency noise limits coherent probing of the atom ensemble to times usually well below 1 s (limiting $\delta\nu$). For times shorter than a measurement cycle, the clock stability is given entirely by the LO.  At longer times, the clock stability is often limited by the Dick effect \cite{Dick1987,SantarelliAudoin1998}: since a measurement cycle includes atom preparation time, only a fraction of the cycle actually probes the atomic transition, i.e.\ measures the LO frequency relative to the transition frequency.  This periodic sampling of the LO aliases higher frequency LO noise, limiting achievable clock stability.  In neutral atom systems, the Dick limit is usually well above fundamental limits such as quantum projection noise \cite{ItanoBergquist1993} and often corresponds closely with experimentally observed clock instability (e.g.\ \cite{UdemDiddams2001,LudlowZelevinsky2008,LemkeLudlow2009}).  Improving LO stability directly reduces the Dick effect both by reducing the frequency noise that is aliased and by enabling longer atomic probing, which reduces the fractional `dead' time between consecutive probe cycles.  Improved LO performance thus plays a critical role in realizing the high stability possible with many optical atomic clocks.  We also note that since stable LOs offer narrower $\delta\nu$, other clock instability limits decrease \cite{ItanoBergquist1993}.  Since instability typically averages down as $1/\sqrt{\tau}$, improving the initial instability greatly reduces the required averaging time, as the square of the improvement.

State-of-the-art laser stabilization usually involves phase-locking a laser source (with electronic feedback) to the mode of a passive, ultra-stable Fabry Perot (FP) cavity (see Figure 1). The FP cavity consists of very high reflectivity mirrors, optically contacted onto a rigid spacer.  In the limit of good S/N and tight phase lock, the length stability of the FP cavity gives the frequency stability of the resulting optical wave.  A fundamental limit that many cavity stabilized lasers have reached is given by Brownian thermal mechanical fluctuations of the FP cavity.  The two cavity mirrors typically dominate the fractional frequency instability limit from thermal noise \cite{NumataKemery2004,NakagawaGretarsson2002}:
\begin{equation}
    \label{eqn:stability2}
    \sigma_{therm} = \sqrt{\textrm{ln}2 \frac{8 k_b T}{\pi^{3/2}}\frac{1-\sigma^2}{Ew_0L^2}\left(\phi_{sub}+\phi_{coat}\frac{2}{\sqrt{\pi}}\frac{1-2\sigma}{1-\sigma}\frac{d}{w_0}\right)}
\end{equation}
Here, $\sigma$, $E$, and $\phi_{sub}$ are Poisson's ratio, Young's modulus, and the mechanical loss for the mirror substrate, $w_0$ is the (laser) beam radius on the mirror, T is the mirror temperature (K), $k_b$ is Boltzmann's constant, and L is the cavity length.  $\phi_{coat}$  and $d$ denote the mechanical loss and thickness of the thin film reflective coating.  The first (second) term in parentheses is the mirror substrate (coating) contribution.  High stability FP cavities are typically made from ultra-low expansion (ULE) glass to reduce cavity length changes with temperature drift around room temperature.  Cavity lengths are often 10--20 cm.  Under such conditions, the lowest thermal noise instability is typically $3$--$10 \times 10^{-16}$, roughly consistent with the best experimentally observed instability (e.g.\ \cite{YoungCruz1999,MilloMagalhaes2009,LudlowHuang2007,WebsterOxborrow2004}).  To reduce thermal noise, choice of mirror substrate material ($E$ and $\phi$), beam waist ($w_0$), cavity length ($L$), and cavity temperature ($T$) can be modified.  Each modification presents different technical difficulties.  Our approach was to fabricate a long cavity featuring a larger beam size and alternative mirror material.

Since thermal expansion of the cavity must be well-controlled, ULE is the best cavity spacer material.  However, $\phi_{sub}$ for fused silica mirror substrates \cite{NumataKemery2004,NotcuttMa2006,MilloMagalhaes2009} is more than ten times smaller than for ULE.  Consequently, the substrate thermal noise term shrinks below that of the thin film reflective coating (Equation \ref{eqn:stability2}, second term), and $\sigma_{therm}$ is here reduced by 1.8.  Now dominated by the thin film coating, $\sigma_{therm}$ can be further reduced with its $1/w_0$ dependence \cite{Levin1998,NakagawaGretarsson2002}.  To do so, we chose a long cavity $L=29$ cm and used mirrors with a larger radius of curvature, $R=1$ m.  The choice of a long cavity also exploits the $1/L$ dependence of $\sigma_{therm}$.  With these cavity parameters, we reduce the thermal-noise-limited fractional-frequency-instability to $1.4 \times 10^{-16}$.

We constructed two cavity systems (Figure 1) and measured the laser coherence properties between them (see Methods). The frequency noise spectrum of one cavity-stabilized laser is shown in Figure 2a.  The noise spectrum approaches the projected thermal noise for Fourier frequencies around 1 Hz; at higher frequencies, the spectrum is approximately white with several spikes attributed to seismic noise for one of the cavities.  In Figure 2b, we show the fractional frequency instability, along with $\sigma_{therm}$.  During typical best performance, at times below 10 seconds we observe an instability as low as $2 \times 10^{-16}$.  Less ideal data sets give a measure closer to $3 \times 10^{-16}$.  At times $>10$ s, laser instability typically increases.  The measured laser power spectrum is shown in Figure 2c, with a 250 mHz linewidth.  To minimize thermal drift of the cavity resonance, we engineer the cavity coefficient of thermal expansion to cross zero near room temperature (Figure 2d, see also Methods).

With the performance highlighted in Figure 2, the cavity-stabilized laser serves well as a low noise LO to probe the narrow clock transition of an optical frequency standard, in this case the $^1S_0$-$^3P_0$ transition of neutral Yb confined in an optical lattice \cite{LemkeLudlow2009}.  The first benefit of the low noise LO is the ability to coherently excite the clock transition for long times (permitting narrower $\delta\nu$).  By probing the transition for 0.9 s and stepping the LO frequency by 0.25 Hz every experimental cycle, we resolve 1 Hz linewidth (FWHM) atomic spectra as in Figure 3a.  With a transition frequency of 518 THz, this corresponds to a line quality factor of $> 5 \times 10^{14}$, the highest achieved for any form of coherent spectroscopy \cite{ChouHume2010b,BoydZelevinsky2006}.  Combined with a good measurement S/N, this makes very low clock instability possible.

To rigorously assess the clock instability (consisting of the LO stabilized to the Yb transition), it must be compared to another standard with even lower instability.  Anticipating previously unobserved stability levels with this individual system, we instead evaluated the noise levels that determine the clock instability.  While the 0.9 s probe time enables very narrow atomic spectra, it leaves the atomic transition lock sensitive to short-time laser frequency excursions which occasionally exceed the $\sim 1$ Hz locking range.  For this reason, we chose to operate the system with a 0.3 s probe time ($\delta\nu \simeq3$ Hz), $40\%$ of the 0.75 s total cycle.  Under these conditions and with the LO frequency noise spectrum from Figure 2a, we calculate the Dick limit to be $1.5 \times 10^{-16}/\sqrt{\tau}$ (Figure 3b, dashed line), an order of magnitude improved over our previous LO \cite{OatesBarber2007}. We tuned the LO frequency on resonance with the atomic transition and used the measured excitation as a frequency discriminator. This measurement is sensitive not only to the LO frequency instability (including the Dick effect) but also to the detection noise of the atom system (photon shot noise, quantum projection noise, technical noise), and thus contains the relevant noise terms for the short to mid-term stability of the optical standard (long-term stability is set by drift of the systematic clock shifts).  The results of this measurement are shown in Figure 3b as blue circles, indicating that the optical clock system supports an instability of $4$--$5 \times 10^{-16}$ after just one measurement cycle.  When we close the loop to lock the LO to the atomic transition, the instability can begin averaging down, suggested by Figure 3b, green squares.  This is the in-loop error signal stability of the LO lock to the atomic transition, and gives a lower limit to the expected clock instability.  At 1 s, we see that our aggressive choice of feedback gain degrades the LO stability.  However, by 2 s we begin averaging down, faster than $1/\sqrt{\tau}$, so that after 20 s of closed-loop operation, we can reach the anticipated $4-5 \times 10^{-16} /\sqrt{\tau}$ indicated by the out-of-loop signal.  Construction of a second Yb optical lattice clock, currently under way, will enable even more direct measure of the improved stability.  We emphasize a key point: by sufficiently reducing the Dick effect, we have more fully enabled the benefits afforded by the S/N of a large atom ensemble to achieve this unprecedented level of stability.

This stability enables high precision frequency measurement at faster speeds than ever before, including characterization of systematic shifts of the Yb clock transition \cite{LemkeLudlow2009}.  Several important systematic shifts can be studied by interleaving the clock operation between two conditions and looking for a frequency shift. The red trace in Figure 3b shows the stability of such a measurement, enabling precision at $< 8 \times 10^{-17}$ at 100 s and averaging down as $1/\sqrt{\tau}$.  Note that this observed instability is consistent with $4-5 \times 10^{-16} /\sqrt{\tau}$ for normal single operation, accounting for the quadrature sum of uncorrelated noise between the two-setting clock operation, as well as the extra cycle time required for dual operation.

Still, these powerful optical clock systems offer more untapped potential.  The Dick limit must be lowered by yet another order of magnitude to reach possible quantum projection noise limits.  Efforts to reduce preparation time in the experimental cycle will help significantly, e.g.\ \cite{WestergaardLodewyck2010}, as can interrogation optimization \cite{Dick1987,SantarelliAudoin1998}.  The effect will continue to get smaller as the LO is further improved, including further reduction of the cavity thermal noise.  Many worthwhile efforts can be made here, focusing on one or more of the critical cavity parameters described above, e.g.\ increasing beam size on the mirror (including higher-order cavity excitation \cite{NotcuttMa2006}) or cryogenic cavity cooling \cite{NumataKemery2004}. To make the cavity effectively longer, a multiple mirror cavity could be used where all but the first and last are folding mirrors.  Each mirror introduces uncorrelated thermal noise, so that a cavity with N mirrors and optical length L between each mirror can have fractional thermal noise scaling as $\sqrt{2N-3}/(N-1)L$ (also a larger average beam size is enabled).  For our $L=29$ cm cavity, with 6 fused silica mirrors, the thermal noise instability limit is reduced to $7 \times 10^{-17}$.

\textbf{Methods}

\textbf{Laser coherence measurements} Several mW of laser light at 578 nm \cite{LemkeLudlow2009} was divided into multiple paths, one to each of the two cavities, and one to the Yb optical lattice apparatus (Figure 1).  Each cavity was enclosed in a vacuum chamber, which was single stage temperature controlled (fluctuations over 24 hours at a few mK).  The cavities were designed with reduced length sensitivity to acceleration induced deformation (similar to mounting designs described in \cite{WebsterOxborrow2007,MilloMagalhaes2009}) and rested on 4 symmetrically placed Viton hemispheres.  The vacuum chamber and optics coupling light to the cavity sat on vibration isolators (one cavity on an active system, the other passive).  Each system resided in different parts of the lab in independent closed acoustic-shielded chambers.  Optical links between the isolated systems and the laser source were made with optical fiber using active phase stabilization \cite{MaJungner1994}.  Free space optical paths were generally in closed boxes to reduce air currents.  The free running 578 nm laser light was locked to Cavity A using Pound-Drever-Hall (PDH) stabilization, by means of fast electronic feedback to an acousto-optic modulator (AOM) common to all optical paths, and slow electronic feedback to a piezo-electric transducer on the laser source.  Thus light incident on both cavities was phase stabilized to Cavity A.  To measure the laser frequency noise spectrum, an additional AOM was used to tune laser light incident on Cavity B into resonance, and the PDH signal of Cavity B served as a frequency discriminator.  To measure the laser frequency stability, the PDH signal from the Cavity B was filtered and fed to an AOM to lock the laser frequency of the second beam to this cavity's resonance.  This AOM frequency thus gave the difference between the two cavities, and was counted to determine the frequency stability.

\textbf{Cavity thermal expansion properties} As described in the main text, each cavity consists of a ULE spacer with fused silica mirror substrates.  Because fused silica mirrors have a much larger CTE ($+500$ ppb/K) than the ULE spacer ($\pm 30$ ppb/K) to which they are contacted, temperature changes create stress which bends the mirror substrate and introduces a larger CTE for the composite-cavity (CTE$_{cav}$) than for ULE alone.  To compensate this problem and achieve a very small CTE$_{cav}$, we exploit ULE's CTE dependence on TiO$_2$ doping level and temperature \cite{Fox2009}.  By obtaining a ULE piece with doping levels corresponding to a room temperature CTE of $-40$ ppb/K, we were able to tune Cavity A temperature so that, combined with the mirror bending effect, we observed a CTE$_{cav}$ zero crossing at 32.2(0.1) C (Figure 2d).  At this temperature (conveniently just above room temperature), CTE$_{cav}<0.2$ ppb/K and the thermally driven drifts are minimized.  For Cavity B, we located a piece of ULE with small, negative CTE at room temperature and, following the proposal of \cite{LegeroKessler2010}, contacted ULE rings to the back side of the fused silica mirrors.  These rings largely prevent the mirror bending from the CTE mismatch with the spacer.  We again see a CTE$_{cav}$ zero crossing, this time at 31.1(0.1) C (Figure 2d).  After characterizing CTE$_{cav}$ for each cavity, to reduce radiative heat transfer, which at first heats only the mirrors (small thermal mass), we installed radiative heat shields (low-emissivity, polished aluminum) between the temperature stabilized vacuum chambers and each FP cavity.

\textbf{Yb optical lattice apparatus} The Yb optical lattice system consists of approximately $10^4$ Yb atoms cooled to a temperature of 10 $\mu$K and trapped in a 1-D optical lattice \cite{LemkeLudlow2009} operating at the magic wavelength where the differential Stark shift on the transition is zero \cite{YeKimble2008,KatoriTakamoto2003}.  The optical lattice facilitates long interrogation of ultracold Yb while eliminating most Doppler and recoil effects.  Excitation of the $m_F=1/2$ state on the $\pi$-clock transition is detected by monitoring the ground and excited state populations after spectroscopic probing and determining the normalized excitation. In addition to the probe time exciting the transition, the experimental cycle typically includes 450 ms dedicated to collecting, cooling, and loading atoms into the optical lattice, as well as state preparation and readout.  Additional noise processes which presently affect the Yb clock instability include atom detection noise (quantum projection noise and technical noise) measured at $\sim 1 \times 10^{-16}$, and uncompensated optical path phase noise on the Yb apparatus measured at $1$--$2 \times 10^{-16}$.

\textbf{Acknowledgements}
The authors gratefully acknowledge optical frequency comb measurement support from S. Diddams, T. Fortier, and M. Kirchner; optical cavity measurement and equipment loan from J. Bergquist, T. Rosenband, and C. Chou; and useful discussions with J. Bergquist and G. Santarelli. We also gratefully acknowledge L. Hollberg for design guidance and useful discussions.

\begin{figure}[c]
\resizebox{12cm}{!}{
\includegraphics[angle=0]{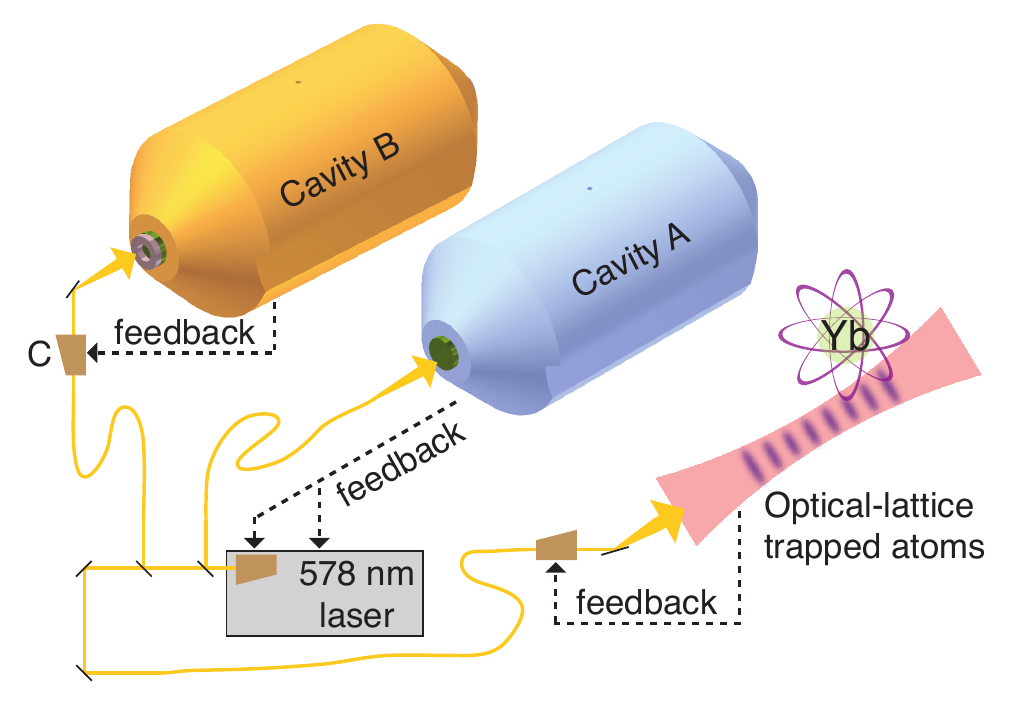}}
\caption{\label{Fig1}(color online) The experimental setup: laser light at 578 nm is incident on two independent, isolated optical cavities.  Each cavity is composed of a rigid ULE spacer with optically bonded fused silica mirror substrates.  Feedback for laser frequency control is usually applied to acousto-optic modulators.  The stabilized light probes the narrow clock transition in an ultracold sample of Yb, confined in a one dimensional optical lattice.}
\end{figure}

\begin{figure}[c]
\resizebox{12cm}{!}{
\includegraphics[angle=0]{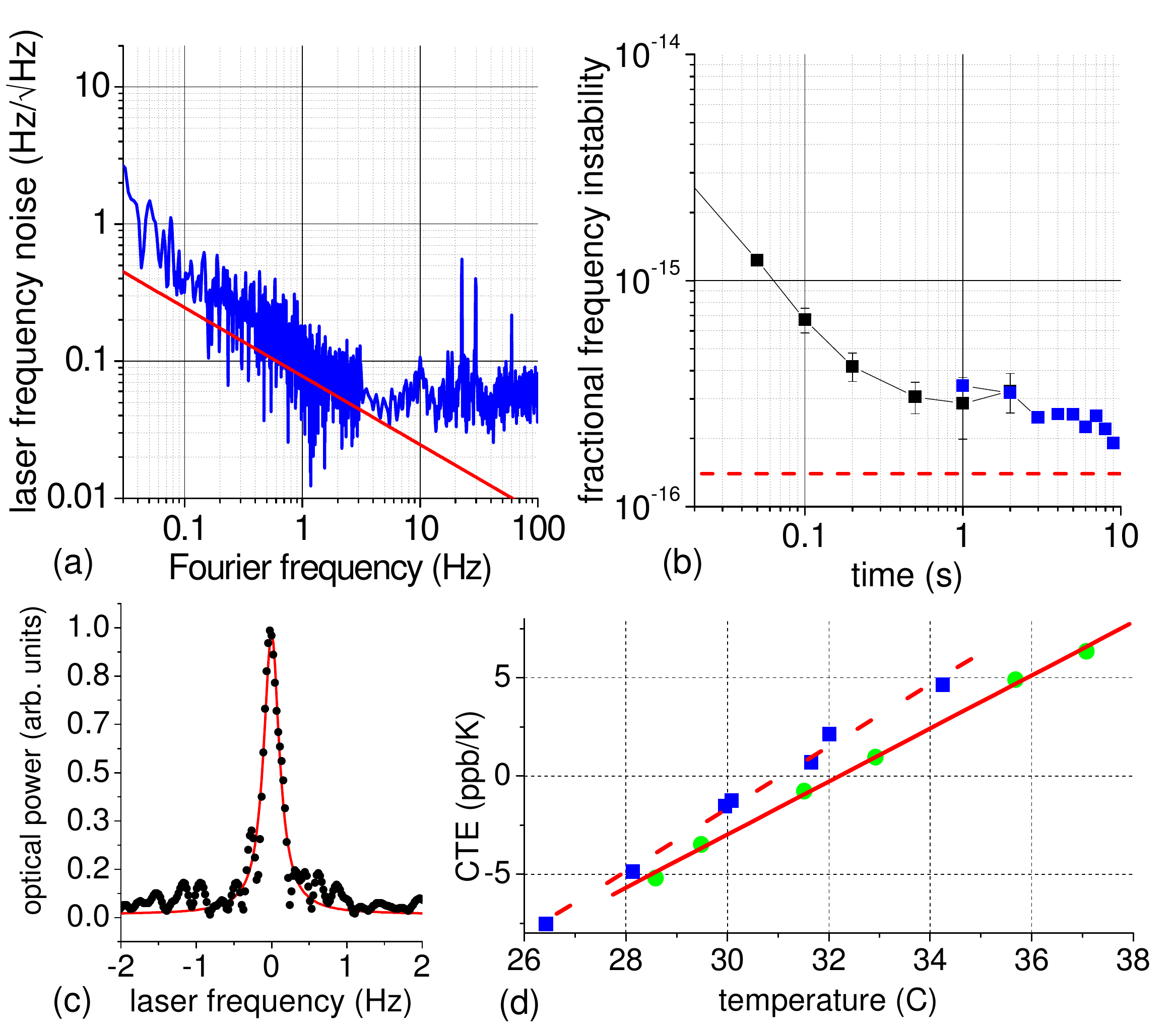}}
\caption{\label{Fig1}(color online) (a) Laser frequency noise spectrum (blue data). The theoretical estimate of the thermal noise contribution, with $\emph{f}^{-1/2}$ dependence, is shown as a solid red line. (b) Fractional frequency instability of one cavity. The blue squares derive from frequency counting with an Agilent 53132 counter \cite{DawkinsMcFerran2007} (with a linear drift removed), while the black squares were measured using the built-in Allan deviation feature of an SRS 620 counter. The red dash curve near a fractional frequency instability of $1.4\times10^{-16}$ denotes the Brownian thermal-noise-limited instability. (c) Laser power spectrum (black dots), measured from the RF source which drives the AOM linking the two stable cavities. The red trace gives a Lorentzian fit with FWHM linewidth of 250 mHz (measurement resolution bandwidth was 85 mHz). (d) Measured CTE$_{cav}$ versus temperature: green dots with a solid linear fit (Cavity A) and blue squares with a dashed linear fit (Cavity B) are measured against a stable optical reference.}
\end{figure}

\begin{figure}[c]
\resizebox{12cm}{!}{
\includegraphics[angle=0]{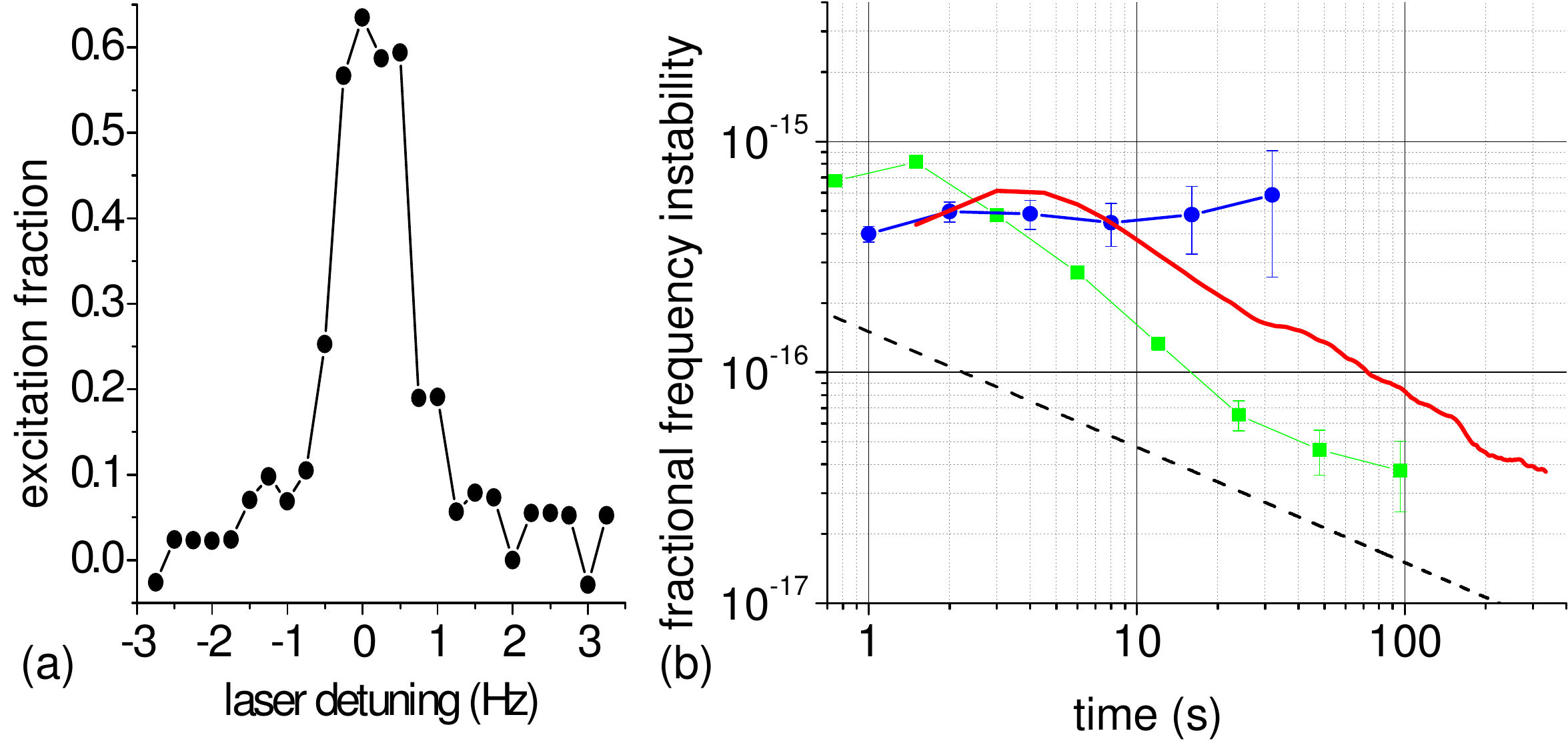}}
\caption{\label{Fig3}(color online) (a) Optical clock
transition spectrum of neutral Yb atoms. The LO frequency is stepped by 0.25 Hz every cycle, with 0.9 s probe time, resolving a spectral linewidth of 1 Hz, FWHM (no averaging) (b) Fractional frequency instability evaluation of the Yb clock using atomic transition. The blue circles correspond to reading off LO frequency using atomic excitation as a (out-of-loop) frequency discriminator.  The green squares are the in-loop stability measured from the error signal used to lock the LO to the atomic transition. The probe pulse duration is 0.3 s. The Dick-effect-limited instability is $1.5 \times 10^{-16}/\sqrt{\tau}$, shown as the black dashed line.  The red curve indicates the stability for interleaved measurements used to assess systematic shifts.}
\end{figure}

\end{document}